\documentclass[aps,rsi,floatfix,twocolumn,superscriptaddress,footinbib,reprint]{revtex4}

\usepackage{amsmath}
\usepackage{amsfonts}
\usepackage{amssymb}
\usepackage{dsfont}
\usepackage{bm}

\usepackage[pdftex]{graphicx}
\usepackage[pdftex]{hyperref}
\usepackage{placeins}
\usepackage{longtable,multirow}
\usepackage{color}

\newcommand{\karman}{von~K\'arm\'an~}

\newcommand{\RX}{\theta_x}
\newcommand{\RY}{\theta_y}
\newcommand{\RZ}{\theta_z}

\newcommand{\lab}{^\mathbb{L}}  
\newcommand{\pa}{^\mathbb{P}} 

\newcommand{\trip}[1]{\underline{#1}}
\renewcommand{\vec}{\bm}

\newcommand{\naxis}{\ensuremath{\hat{\bm n}} }

\newcommand{\matrize}[1] {\underline{\underline{\mathbf{#1}}}}

\newcommand{\degree}{\ensuremath{^\circ} }
\DeclareMathOperator{\Tr}{Tr}
\DeclareMathOperator{\acos}{acos}
\DeclareMathOperator{\asin}{asin}
\DeclareMathOperator{\atan}{atan}
\DeclareMathOperator{\sign}{sign}

\providecommand{\abs}[1]{\left\lvert#1\right\rvert} 
\providecommand{\norm}[1]{\left\lVert#1\right\rVert}

\DeclareMathOperator{\erf}{erf}

\newcommand{\one}{\ensuremath{\mathds{1}}}

\newcommand{\eg}{\emph{e.g.~}} 
\newcommand{\ie}{\emph{i.e.~}} 
\newcommand{\fig}[1]{Fig.~\ref{#1}} 
\newcommand{\gleich}[1]{Eq.~\eqref{#1}} 

\DeclareMathOperator{\cm}{cm}

\definecolor{darkblue}{rgb}{0,0,0.5}
\definecolor{darkgreen}{rgb}{0,0.4,0}
\definecolor{darkred}{rgb}{0.4,0,0}
\newcommand{\Cred}[1]{{\color{red} #1}}
\newcommand{\Cgreen}[1]{{\color{darkgreen} #1}}

\newcommand{\Cblue}[1]{{\color{blue} #1}}

\hypersetup{pdftex=true, colorlinks=true, breaklinks=true, linkcolor=darkblue, menucolor=darkblue,  urlcolor=darkblue}

\begin{document}

\title{Tracking the dynamics of translation and absolute orientation of a sphere in a turbulent flow}
\author{Robert Zimmermann}
\affiliation{International Collaboration for Turbulence Research}
\affiliation{Laboratoire de Physique, CNRS UMR 5672, Ecole Normale Sup\'erieure de Lyon, Lyon, F-69007 France}

\author{Yoann Gasteuil}
\affiliation{International Collaboration for Turbulence Research}
\affiliation{Laboratoire de Physique, CNRS UMR 5672, Ecole Normale Sup\'erieure de Lyon, Lyon, F-69007 France}

\author{Mickael Bourgoin}
\affiliation{International Collaboration for Turbulence Research}
\affiliation{Laboratoire des \'Ecoulements G\'eophysiques et Industriels, CNRS UJF INPG, F-38041 France}

\author{Romain Volk}
\affiliation{International Collaboration for Turbulence Research}
\affiliation{Laboratoire de Physique, CNRS UMR 5672, Ecole Normale Sup\'erieure de Lyon, Lyon, F-69007 France}

\author{Alain Pumir}
\affiliation{International Collaboration for Turbulence Research}
\affiliation{Laboratoire de Physique, CNRS UMR 5672, Ecole Normale Sup\'erieure de Lyon, Lyon, F-69007 France}

\author{Jean-Fran\c{c}ois Pinton}
\affiliation{International Collaboration for Turbulence Research}
\affiliation{Laboratoire de Physique, CNRS UMR 5672, Ecole Normale Sup\'erieure de Lyon, Lyon, F-69007 France}

\date{\today}

\begin{abstract}
We study the 6-dimensional dynamics -- position and orientation -- of a large sphere advected by a turbulent flow. The movement of the sphere is recorded with 2 high-speed cameras. Its orientation is tracked using a novel, efficient algorithm; it is based on the identification of possible orientation `candidates' at each time step,  with the dynamics later obtained from maximization of a likelihood function. Analysis of the resulting linear and angular velocities and accelerations reveal a surprising intermittency for an object whose size lies in the integral range, close to the integral scale of the underlying turbulent flow.
\end{abstract}

\maketitle

\section{Introduction}
The advent of resolved Lagrangian measurements has helped understand the dynamics of turbulence from the point of view of fluid particles~\cite{annRev:LagProps}. In the experiments, solid \emph{tracers} are followed in lieu of   fluid particles, which naturally raises the question of the understanding of the dynamics of \emph{finite size} objects in turbulent flows. It is a subclass of the issue of the dynamics of \emph{inertial particles}, \ie particles who have inertia with respect to the fluid motions, either because their density differs from that of the fluid or because their spatial extent cannot be ignored. If the particles are quite small compared to the smallest fluid motion (the Kolmogorov dissipative length scale $\eta$), arguments show that they behave as tracers of fluid motions. Observations have revealed a very intense intermittency in the motion of fluid tracers~\cite{nature:accEB,mordant2001measurement}. They experience very strong accelerations, with a probability distribution which displays stretched exponential tails~\cite{mordant:kernel}.

When the diameter $D$ of the advected particles is of the order of, or larger than  $\eta$, their equation of motion is not known (see, however~\cite{auton:1988,lovalenti:1993,loth:2009}). We restrict our discussion to neutrally buoyant spheres. Several recent studies~\cite{qureshi2007turbulent,qureshi2008acceleration,brown:2009,volk2010dynamics}  have shown that the acceleration statistics of such inertial particles does not gently reduce to a Gaussian behavior as their diameter increases. It is an important feature because the characterization of forces acting on an object advected by a turbulent flow has many applications in engineering (from mixing issues in industrial processes to dispersion in the oceans or in the atmosphere). 


The study reported here takes a leap forward in size and considers the motion of a neutrally buoyant sphere with diameter $D$ of the order of the integral scale $L_\text{int}$ (the scale at which energy is fed into the flow). In addition, we aim at resolving the six degrees of freedom of the particle dynamics, \ie the goal is to obtain a tracking in time and space of the particle's linear and angular motions. This allows the study of the forces and torques acting on a (large) inertial particle.

The tracking of the particle position in space can be carried out by using methods already developed and successfully tested for small particles~\cite{book:expFluids}. In comparison, following the orientation of the particle is much more challenging, both because of the specifics of angular variables, and of specific algorithmic requirements. Previous studies on tracking the orientation have focused on measuring one component of the angular velocity. The particles~\cite{Frish:1981fk} used for this purpose are transparent, and contain an embedded mirror and a diameter of less than 50$\mu$m, which is of the order of  the Kolmogorov length scale, $\eta$. In the experiments reported in~\cite{Frish:1981fk}, the angular velocity at  a point neither the translation nor the angular motion could be tracked for very long. The principle used here is completely different: it consists simply in painting the particle with a suitable layout, and in retrieving its orientation. For algorithmic efficiency (and robustness) this is not done step by step but for the entire trajectory using a global path extraction.

{The text below is organized as follows: we first present the experimental setup and recall some important features of the orientation algebra in 3D.  We then describe how the particle images are extracted from the movie images, and compared to synthetic images with arbitrary orientations. Possible candidates are identified and then assembled into an orientation time series using a global maximization of a likelihood function.  Finally, we present some results concerning the particle dynamics.}

\section{Basics}
\subsection{Experimental Setup}
\begin{figure}[tb] 
   \centering
   \includegraphics[width=\columnwidth]{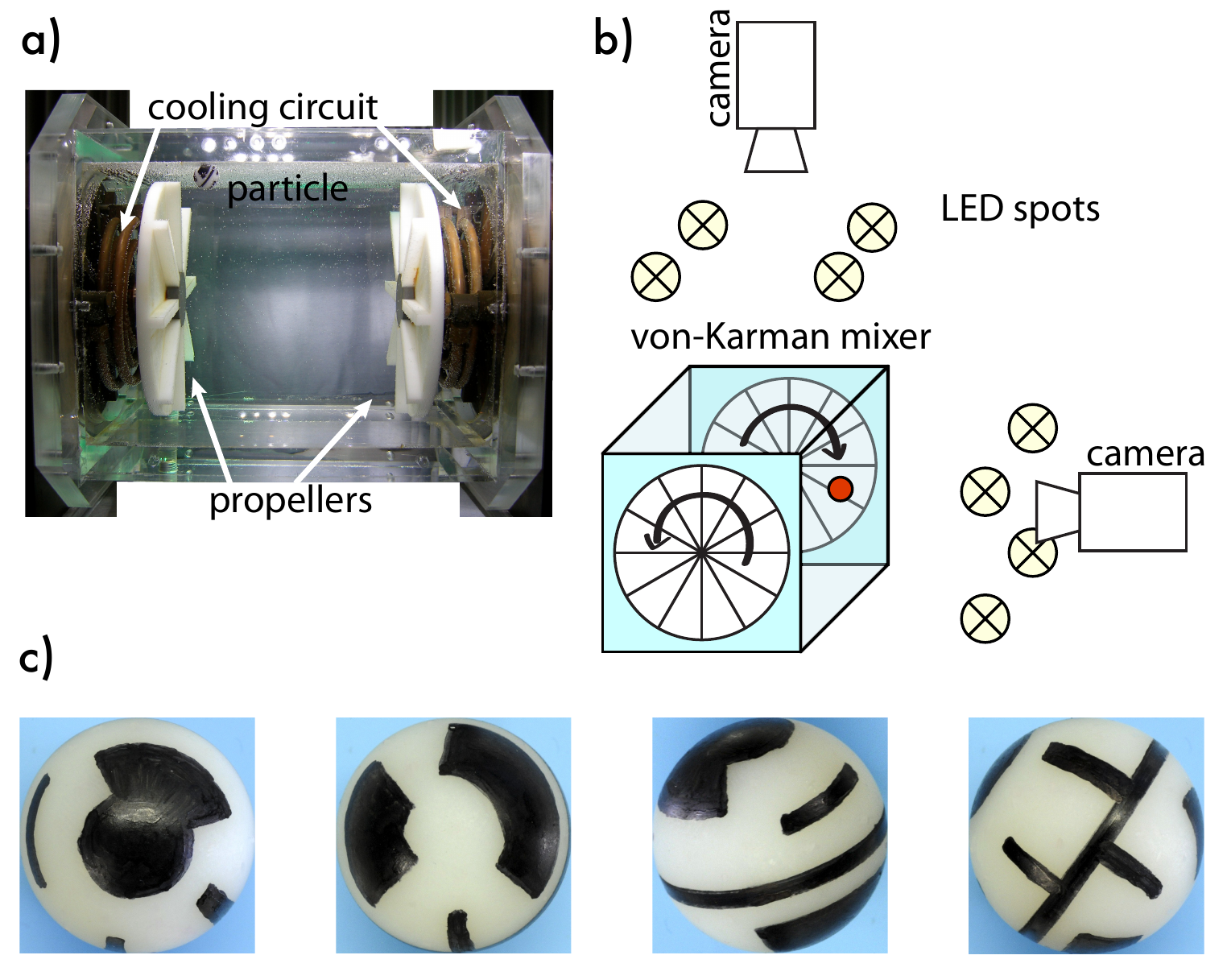} 
   \caption{Sketch of the experimental setup: a) image of the \karman mixer; b) sketch of the camera arrangement; c) textured sphere for different orientations.}
   \label{fig:KLAC}
\end{figure}

A turbulent flow is generated in the gap between 2 counter-rotating impellers of radius $R=10$~cm fitted with straight blades $1$~cm in height. The flow domain in between the impeller has characteristic lengths $H = 2R = 20$~cm and the working fluid is a water-glycerol mixture, whose density can be finely tuned.
In order to be able to perform direct optical measurements, the container is build with flat Plexiglas (Poly[methyl methacrylate]) side walls, so that the cross section of the vessel is square.   This type of \karman swirling flow has been used extensively in the past for the study of fully developed turbulence~\cite{annRev:LagProps}; its local characteristics approximate homogeneous turbulence in its center, although it is known to have a large scale anisotropy~\cite{ouellette:2006, Monchaux:2006fj}.  A sketch of the setup is provided in \fig{fig:KLAC}  -- further details about the flow turbulence are given later in section~\ref{sec:physik}.

A white, PolyAmid sphere  with diameter $D=18$~mm (accuracy $0.01$ mm, Marteau \& Lemari\'e, France) moves and rotates in the turbulent flow. It is neutrally buoyant in the fluid -- whose  density is adjusted to that of the particle $\rho_p = 1.14 \; {\rm g}.{\cm^{-3}}$ by  addition of glycerol to water. The density mismatch, measured from sedimentation speeds, is found to be less than $\Delta \rho \,/ \rho = 10^{-4}$.  The particle is textured black and white by hand using either black nail polish or a black-ink permanent marker.  Its motion is tracked using 2 high-speed video cameras (Phantom V12, Vision Research) which record synchronously 2 views at approximately 90 degree. The flow is illuminated by high power LEDs and sequences of 8~bit gray scale images are recorded at a rate of 600 frames per second.

Both cameras observe the measurement region with a resolution of  $650\times 650$ pixels, covering a volume of $15\times 15\times 15 \; [\cm^3]$. Hence,  the particle  diameter  is $70-90$ pixels.
In the choice of the particle texture, several features have to be considered:\\
- a single view should correspond to a unique orientation. \\
- illumination inhomogeneities may cause regions to look similar in the camera images.  Optically resembling views should correspond to clearly distinct orientations. \\
- the cameras are grayscale so the texture has to be black and white.\\
- the number of black and white pixel should be approximately the same in every possible view.

In our configuration, the camera can store on the order of $15,000$ frames in on-board memory, thus limiting the duration of continuous tracks. The movies are downloaded to a PC, waiting to be processed. The processing is done on a gaming PC with a state of the art graphics card. Algorithm development and code test is done on an Apple Macbook Pro.  The code is written in Matlab 2009a using the image and signal processing toolboxes as well as the Psychtoolbox extension\cite{ptb1,ptb2} which provide OpenGL wrappers for Matlab.

\subsection{Angular Variables}
The parametrization of an angular position in 3D space causes a number of difficulties which are briefly addressed in this section (see \eg \cite{goldstein,aircraft,mathworld} for a more complete presentation). One of them is caused by the degeneracy of the axes of rotation for certain orientations (the `gimbal lock' problem). Another is the choice of a suitable measure of distance between two orientations.

\subsubsection{Describing Orientations}
\begin{figure}[htbp] 
   \centering
   \includegraphics[width=\columnwidth]{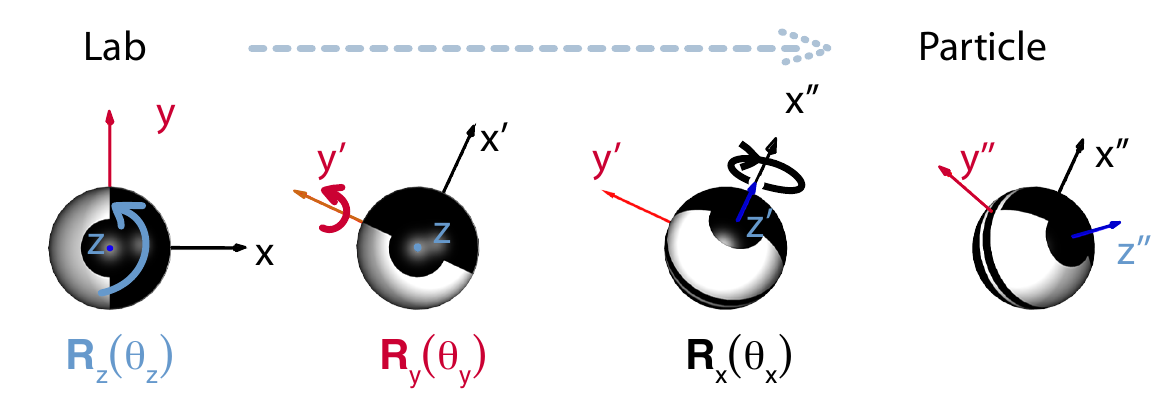} 
   \caption{Tait-Bryan rotation sequence describing the sphere's orientation.}
   \label{fig:rotSeq}
\end{figure}

As stated by the Euler rotation theorem, 3 parameters are needed to describe any rotation in 3D. We use here Euler angles with the Tait-Bryan convention as shown in \fig{fig:rotSeq}. In the transformation from Lab to Particle coordinate system (CS), we first apply a rotation around the $z-$axis of angle $\RZ$, followed by a rotation around the intermediate $y-$axis of angle $\RY$ and last a rotation of angle $\RX$ around the new $x-$axis. The rotations work on the object using a right handed coordinate system and right handed direction of rotation. We will denote an orientation triplet by an underscore, \eg $\trip\theta$, in order to distinguish them from vectors (which are typeset in bold font, \eg~$\vec\omega$).

The orientation of the object is fully described by an orthogonal $3\times 3$ matrix $\matrize R$, obtained from the composition of the 3 elementary rotations: 
\begin{equation}\begin{split}
&\matrize R(\RX,\RY,\RZ)=\matrize R_x(\RX)\matrize R_y(\RY)\matrize R_z(\RZ)=\\
&\begin {bmatrix}
c \RY \, c  \RZ 			& -c \RY  \,s  \RZ   			&{s  \RY}  \\
s  \RX \, s  \RY  \,c  \RZ  +c \RX  \,s  \RZ  		&-s  \RX  \,s  \RY \, s  \RZ  +c  \RX  \,c  \RZ 	&{-s  \RX  \, c  \RY}  \\
-c  \RX  \,s  \RY \, c  \RZ  +s  \RX \, s  \RZ  	&c  \RX  \,s  \RY \, s  \RZ  +s  \RX  \,c  \RZ  	&{c  \RX  \,c  \RY}  
\end{bmatrix}
\end{split}
\end{equation}
with $c\cdot=\cos\left(\cdot\right)$ and $s\cdot=\sin\left(\cdot\right)$. 
Consequently, from any rotation matrix the 3 Euler angles can be extracted using
\begin{equation}\begin{split}
\trip\theta=\{\RX,\RY,\RZ\}=
\begin{pmatrix}
{\rm atan2}({-\matrize{R}_{12},\matrize{R}_{11}})\\
 \asin({\matrize{R}_{13}})\\
 {\rm atan2}({-\matrize{R}_{23},\matrize{R}_{33}})\end{pmatrix} \ ,
\end{split}\end{equation}
enforcing $\RX,\RZ \in [0, 2\pi[$ and $\RY\in [-\pi/2, \pi/2]$. However this choice is not unique because there is a second triplet with
 $\matrize R \left(\RX+\pi,\sign(\RY)\cdot \pi- \RY, \RZ+\pi\right)=\matrize R\left(\RX,\RY,\RZ\right)$.   Needless to say, multiples of $2\pi$ can be added to each angle. An important practical consequence is that even for small changes in orientation the difference between 2 Euler angle triplets, $\trip\theta_1$ and $\trip\theta_2$, has formally 4 possible results. 
  
The curvilinear coordinate $\trip\theta$ is related to the angular velocity, $\vec \omega\pa$, (in the particle frame) by 
\begin{equation}
\begin{split}
\vec\omega\pa\big(\trip\theta(t)\big)&= 
\begin{bmatrix}1 & 0 & s\RY \\
 0 & c\RX & -s\RX\,c\RY \\
 0 & s\RX & ~c\RX\,c\RY\end{bmatrix}\cdot
 \frac{d}{dt}\begin{pmatrix}\RX\\
  \RY \\
   \RZ \end{pmatrix}
\\ &= \matrize H(\RX,\RY) \cdot \frac{d}{dt}\begin{pmatrix}\RX\\
  \RY \\
   \RZ \end{pmatrix} \ .
\end{split}
\label{eq:omegaalpha}
\end{equation} 
For $\cos\left(\RY\right)\approx 0$, the determinant of the matrix $\matrize H$, $\det(\matrize H)$, vanishes and its inverse is not defined. In other words, finite body rotations need infinite change in the Euler angles.
This singularity is called a \emph{gimbal lock} and is a well-known problem in robotics and aerospace engineering. 
Geometrically, the second rotation turns the first axis parallel to the third axis of rotation, and the rotation loses 2 degrees of freedom. 
Unfortunately gimbal locks cannot be avoided by a wise choice of representation. \\

One then needs to define a distance between 2 arbitrary orientations, immune to this type of singularity. 
A natural   distance between two arbitrary orientation matrixes, $\matrize A$ and $\matrize B$, is
\begin{equation}\begin{split}
&\Tr \left( \left(\matrize A - \matrize B\right) \left(\matrize A -\matrize B\right)^T\right)=
6 -2\Tr\left(\matrize A \matrize B^T\right)\\
&= 4\big(1- \cos\left(\phi\right)\big)
\end{split}\label{eq:trace}\end{equation}
using the $\matrize A \,\matrize A^T=\matrize B \,\matrize B^T=\one$ and that $\matrize A \matrize B^T$ is a rotation matrix with the eigenvalues $1, e^{i\phi},e^{-i\phi}$. The distance is thus a growing function of 
$\phi$. We measure here the distance between two rotation matrices by:
\begin{equation}
d \left( \matrize A , \matrize B \right) \equiv \acos \left( \frac{1}{2} \left[ \Tr \left( \matrize A \, \matrize B^T \right) -1 \right] \right)
\end{equation} 
Because it works directly on the orientation matrices it is neither sensitive to gimbal locks nor to the choice of the representation and thus an important tool in our algorithm. It should be noted that $d(\matrize A,\matrize B)$ is the angle of the rotation which turned the orientation from $\matrize A$ to $\matrize B$. \\

In the search of the particle orientation, one last inconvenience of Euler angles is that they are not locally orthogonal, in the sense that
\begin{equation}\begin{split}
d\big(\{\RX,\RY,\RZ\},\{\RX+\Delta\RX,\RY+\Delta\RY,\RZ+\Delta\RZ\}\big)^2\approx\\ \Delta\RX^2+\Delta\RY^2+\Delta\RZ^2+2\Delta\RX \cdot \Delta\RZ\cdot \sin(\RY)
\end{split}\end{equation} 
for $\Delta$ small. As a consequence, a uniform spacing of the Euler angles in $\RX,\RY,\RZ$ does not sample the space of possible orientations in an optimal way. In particular near gimbal locks, the sampling rate would be higher at no higher accuracy. The so-called Lattman angles~\cite{lattman1972optimal}
\begin{equation}
\{\theta_+,\theta,\theta_-\}\equiv\{\RX +\RZ, \RY,\RX-\RZ\}
\end{equation} 
fulfill  local orthogonality since they verify
\begin{equation}\begin{split}
d\big(\{\theta_+,\theta,\theta_-\},\{\theta_++\Delta\theta_+, \theta +\Delta \theta,\theta_-+\Delta\theta_-\}\big)^2\approx\\
\Delta\theta_+^2(1+\sin \theta)/2 +\Delta\theta^2+ \Delta\theta_-^2(1-\sin \theta)/2 \ . \label{eq:chiLattman}
\end{split}
\end{equation}
As they are locally orthogonal, it is sufficient for sampling purposes to keep $\Delta\theta_+^2(1+\sin \theta)/2$,  $\Delta\theta^2$ and $\Delta\theta_-^2(1-\sin \theta)/2$ constant. After a constant sampling of $N$ values of $\theta$  with $\Delta_\text{Latt}\equiv\Delta\theta={\pi}/({N-1})$, the stepping in $\theta_+$ and $\theta_-$ can be computed with 
$\Delta\theta_+(\theta)={\Delta_\text{Latt}}/{\sin\left(\frac{\theta}{2} + \frac{\pi}{2}\right) }$ and $\Delta\theta_-(\theta)={\Delta_\text{Latt}}/{\sin\left(\frac{\pi}{2}-\frac{\theta}{2} \right) } $. It should be emphasized, that $\theta_-\in[0, 2\pi[$ whereas $\theta_+\in[0, 4\pi[$. Lattman angles enable us to sample the set of orientations in an optimal way.\\

Finally, in several instances it is convenient to describe a rotation by the direction of an axis \naxis about which the systems is rotated by an amount $\phi$. The corresponding rotation matrix can be computed using the Rodrigues Formula~\cite{goldstein,mathworld} 
\begin{equation}
\begin{split}
&\matrize R\left( \naxis,\phi \right)=\\
&\begin{bmatrix} 
c\phi+n_x^2 A 	& -n_z s\phi +n_x n_y A		&	n_y s\phi+n_x n_z A\\
n_z s\phi+n_x n_y A	& c\phi+n_y^2 A			&	-n_x s\phi+n_y n_z A\\
-n_y s\phi+n_x n_z A	& n_x s\phi+n_y n_z A	&	c\phi+n_z^2 A
 \end {bmatrix}\\
 &\text{with~} A=(1-c\phi) \ .
\end{split}\label{eq:rodrigues}
\end{equation}
\gleich{eq:rodrigues} also allows us to extract the axis, \naxis, and the angle, $\phi$ from any arbitrary rotation matrix. As a result, changing the coordinate system or changing the representation of rotation can be done by expressing the orientation in its  matrix form, applying the transformation which changes the CS and extracting the desired representation.\\

\subsubsection{Angular Velocity and Acceleration}
Angular velocity and acceleration are often obtained by direct differentiation of a time-series of Euler angles, \eg using \gleich{eq:omegaalpha}. However, it is possible to obtain the angular velocity in the particle frame directly from the matrices. This technique is not sensitive to Gimbal locks because of the uniqueness of the orientation matrices. 

Let $\vec e^{\mathbb{P},k}_{x,y,z}$ be the particle CS at time step $k$, whereas the fixed lab CS is $\vec e^{\mathbb{L}}_{x,y,z}$.
For two time-steps, $k$ and $k+m$, we know the corresponding orientation {matrices} which rotate the particle:
 \begin{align*}
&{\matrize R (\trip\theta_k)}~	&:\qquad  \vec e_{x,y,z}\lab & \quad \xrightarrow{\matrize R (\trip\theta_k)} 	\qquad	&~ &\vec e^{\mathbb{P},k}_{x,y,z}\\
&{\matrize R (\trip\theta_{k+m})}~&:\qquad  \vec e_{x,y,z}\lab & \quad \xrightarrow{\matrize R (\trip\theta_{k+m})} \qquad	&~ &\vec e^{\mathbb{P},k+m}_{x,y,z}\\
&{\matrize T}~&:\qquad  \vec e^{\mathbb{P},k}_{x,y,z} & \quad \xrightarrow{\matrize R (\trip\theta_{k+m})~\matrize R (\trip\theta_k)^T} \qquad	&~ &\vec e^{\mathbb{P},k+m}_{x,y,z}
\end{align*} %
 in which the matrix $\matrize T$ is the change in orientation, in other words the matrix representation of the discrete angular velocity (for a given time difference). The change is with respect to the particle CS at time $k$: $\vec e^{\mathbb{P},k}_{x,y,z}$.
 $\matrize T$ expressed in the axis-angle convention (see \gleich{eq:rodrigues}) returns a  direction vector, \naxis, of length unity and an angle, $\Delta\phi$ (meaning that between times $k$ and $k+m$ the particles has rotated an angle $\Delta\phi$ around the vector \naxis). The time difference, $\Delta t$, between the steps is a function of $m$. Therefore an estimator of angular velocity is
\begin{equation}\begin{split}
\vec \omega^{\mathbb{P}}\big(t(k)\big)=
  \frac{\Delta\phi}{\Delta t(m)} \left(	n_x \cdot \vec e_x^{\mathbb{P},k} + n_y \cdot \vec e_y^{\mathbb{P},k} +	n_z \cdot \vec e_z^{\mathbb{P},k} \right)
\end{split}\end{equation}
Averaging $\naxis \frac{\Delta\phi}{\Delta t}$ over several separations, $m$, returns the angular velocity in the particle frame without  a prior  unwrapping nor problems near gimbal locks. The angular velocity with respect to the lab CS is defined as
\begin{equation}\begin{split}
\vec \omega^{\mathbb{L}}\big(t(k)\big)=\matrize R\left( \trip\theta_k\right) \vec \omega^{\mathbb{P}}\big(t(k)\big)
\end{split}\end{equation}

The angular acceleration in either particle or lab frame is defined as
\begin{equation}\begin{split}
\vec\alpha^{\mathbb{L/P}}=\frac{d}{dt}\vec\omega^{\mathbb{L/P}}
\end{split}\end{equation}
In practice, it is obtained from a convolution of the angular velocity time series with the derivative of a gaussian kernel. This technique has proved to be efficient in removing noise~\cite{mordant:kernel}.

\section{Tracking}
\subsection{Position}\label{ss:pos}
Although the identification of a large sphere from the camera images causes no particular conceptual difficulty, the fact that the sphere is \emph{textured} raises some practical issues. A simple thresholding returns only either the white or the black part of the particle. Reflections from the impellers continuously  change the background,  and small impurities in the flow and possible bubbles add sharp gradient noise to the images. Furthermore, the illumination of the flow is not perfectly uniform, and thus, shadows as well as reflections occur.

For each movie and for each camera, we compute the background view as the average of an equally distributed subset of its images. For each frame we then subtract the background and perform a \emph{Difference of Gaussians} blob detection. The threshold is adjusted by hand for each camera and light arrangement.  Matlab's Image-processing toolbox is used to identify blobs with a round shape and a diameter close to that of the particle.  Shadows, bubbles, and reflections might be found during blob detection because of their sharp separation from the background, but they are of uniform texture and hence characterized by a small value of the variance of light intensity across the blob. The blob with highest variance and closest resemblance to {a} sphere is considered to be the particle. The precise position of the particle is refined using a \emph{Canny edge} detection in a tight region around the blob. For each time step we record the position,  $(x,y)$, of the particle on the image in pixels plus its diameter, $2\,r$,  and the deviation from the spherical shape as an error estimator. Since only one particle is placed into the flow,  the track assembly is straight forward. The algorithm may temporarily loose the particle for short times (because of bad light reflection, blurs, ...); this is compensated by the large oversampling and gaps of less than 5 frames are interpolated to obtain longer tracks. Outliers are be identified using a least square spline and replaced by an interpolation.

Tsai's camera model and  calibration technique\cite{tsai:vcc} is used to project the 2D positions into 3D. The calibration of the cameras contains the position of the camera plus its rotation with respect to the lab CS, which is needed later for the orientation processing.

\subsection{Orientation}
The algorithm used to process the camera images and obtain a time series of orientations (and angular velocities) can be split into 3 parts: (i) by comparison of the sphere's picture with synthetic images, the algorithm identifies a set of possible orientations; (ii) from the set of possible candidates at successive instants, a \emph{Flow} algorithm identifies a likely time series; (iii) a post-treatment adjusts remaining ambiguities. These steps are described in details in this section.
\begin{figure}[b!] 
   \centering
   \includegraphics[width=\columnwidth]{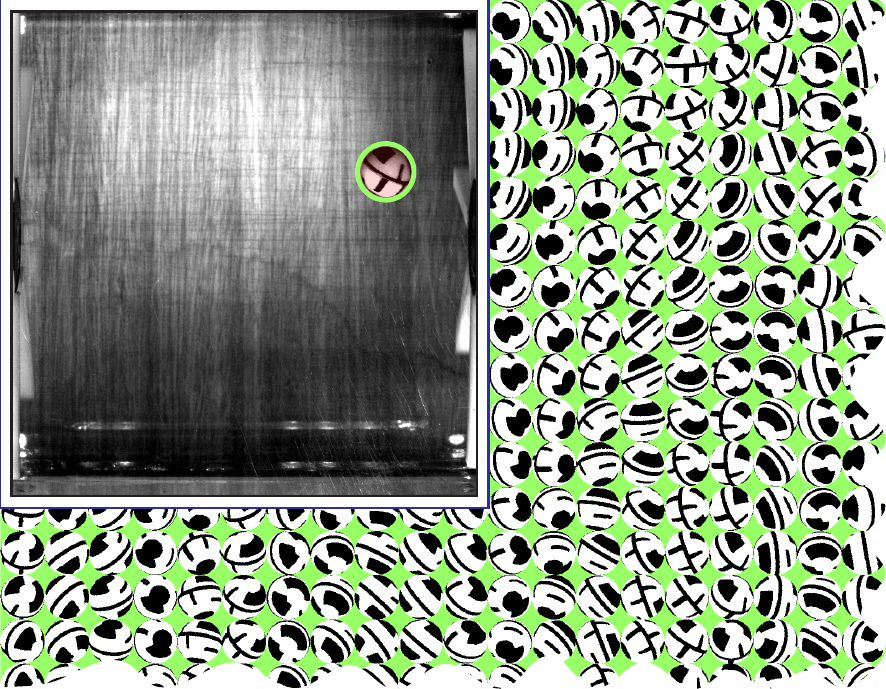} 
   \caption{Synthetic 2D projections of the particle for a range of orientations, using OpenGL. A camera image of the moving particle is shown in the upper left corner (contrast enhanced; note the driving disks on either side). }
   \label{fig:render}
\end{figure}

\begin{figure*}[ht!] 
   \centering
   \includegraphics[width=\textwidth]{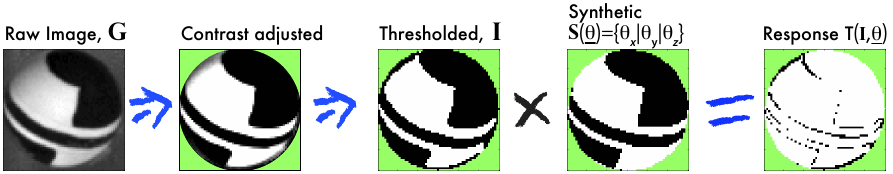} 
   \caption{Texture extraction and comparison with a synthetic image. The resemblance between the image $\matrize I$ and the synthetic projection  $\matrize S$ at angle $\trip\theta$ is estimated using \gleich{eq:trust}. }
   \label{fig:3color}
\end{figure*}

\subsubsection{Candidate Finding}

\paragraph{Synthetic images.}
A first step is to obtain a 2D projection,  $\matrize S(\trip\theta)$, of a sphere with known texture and size at an arbitrary orientation, $\trip\theta$.  This rendering is achieved using OpenGL, \emph{via} the Psychtoolbox extensions for Matlab -- for a disk image of about 60 pixels, the algorithm can render several thousand orientations per second (see \fig{fig:render} for an illustration).\\

\paragraph{Texture extraction.}
Once the particle position and diameter are known, one extracts a disk subset of the image, centered on the particle, $\matrize G$. In a first step the contrast is adjusted such that the global histogram of intensity contains at least $b$ percent of black and $w$ percent of white pixels (the algorithm only takes into account the disk / particle region in $\matrize G$). The adjustable parameters $b$, $w$ are  fixed  to $b = w \sim 30\%$ which is the minimum amount of black/white pixel in an arbitrary orientation.
 In a second step,  the image is thresholded using Otsu's method\cite{Otsu:1979fk} for the global histogram as well as for 2 moving regions.  The thresholded image, $\matrize I$, is adjusted such that pixels outside the particle / disk are set to $0$ whereas black is $-1$ and white $+1$.  These steps are shown in~\fig{fig:3color}.   \\

\paragraph{Comparison, possible orientations.}\label{pp:cands}
The image $\matrize I$ (with diameter $2\,r$) obtained as above is ready for comparison with synthetic images.  The resemblance to a rendered image $\matrize S(\trip\theta)$ with orientation $\trip\theta$ is estimated using the projection 
\begin{equation}\begin{split}
T(\matrize I,\trip\theta)=\frac{1}{2}+\frac{1}{2 \pi r^2}\sum_{i}\sum_{j}\matrize I_{i,j}\cdot \matrize S_{i,j}(\trip\theta) \ ,
\label{eq:trust}
\end{split}\end{equation}
which is ratio of the number of correct pixels to the total number of pixels.

At this point we note that the computational cost of directly comparing  an image $\matrize I$ to synthetic ones $\matrize S(\trip\theta)$ covering the set of possible orientation $\{ \trip\theta \}$ scales roughly as $\Delta_\text{Latt}^{-3}$, where $\Delta_\text{Latt}$ is the grid spacing in the orientation space. There is also the additional difficulty that the particle apparent diameter changes slightly as the sphere moves in the flows. For efficiency and physical correctness, we use the following strategy:  instead of finding at any time step the best images, we identify a set of possible candidates for all time steps and then extract globally the time series of orientations. 

First we render images, $\matrize S(\{\trip\theta_\text{coarse}\})$, covering \emph{all} possible orientations with a coarse grid -- in practice $\Delta_\text{Latt} \approx 12^\circ$. Lattman angles are  locally orthogonal and thus more efficient in creating such grids. The size of the rendered images is fixed to approximately one half of the particle real diameter. Since their size does not change, these images are kept in the computer memory and do not need to be recomputed for every new image.

The thresholded particle image, $\matrize I$, is then resized to the size of the renderings, $\matrize I_\text{coarse}$, and compared to all synthetic images, $\matrize S\left(\left\{\trip\theta_\text{coarse}\right\}\right)$ as shown in \fig{fig:render} using \gleich{eq:trust}. All angles with $\trip\theta>\max(T(\matrize I_\text{coarse},\{\trip\theta_\text{coarse}\}))-\delta_\text{coarse}$ are considered to be possible orientations (PO). Here $\delta_\text{coarse}$  is an arbitrary thresholding value, with inspection showing that a value equal to 0.1 gives good results. 

\begin{figure}[htb] 
   \centering
   \includegraphics[width=\columnwidth]{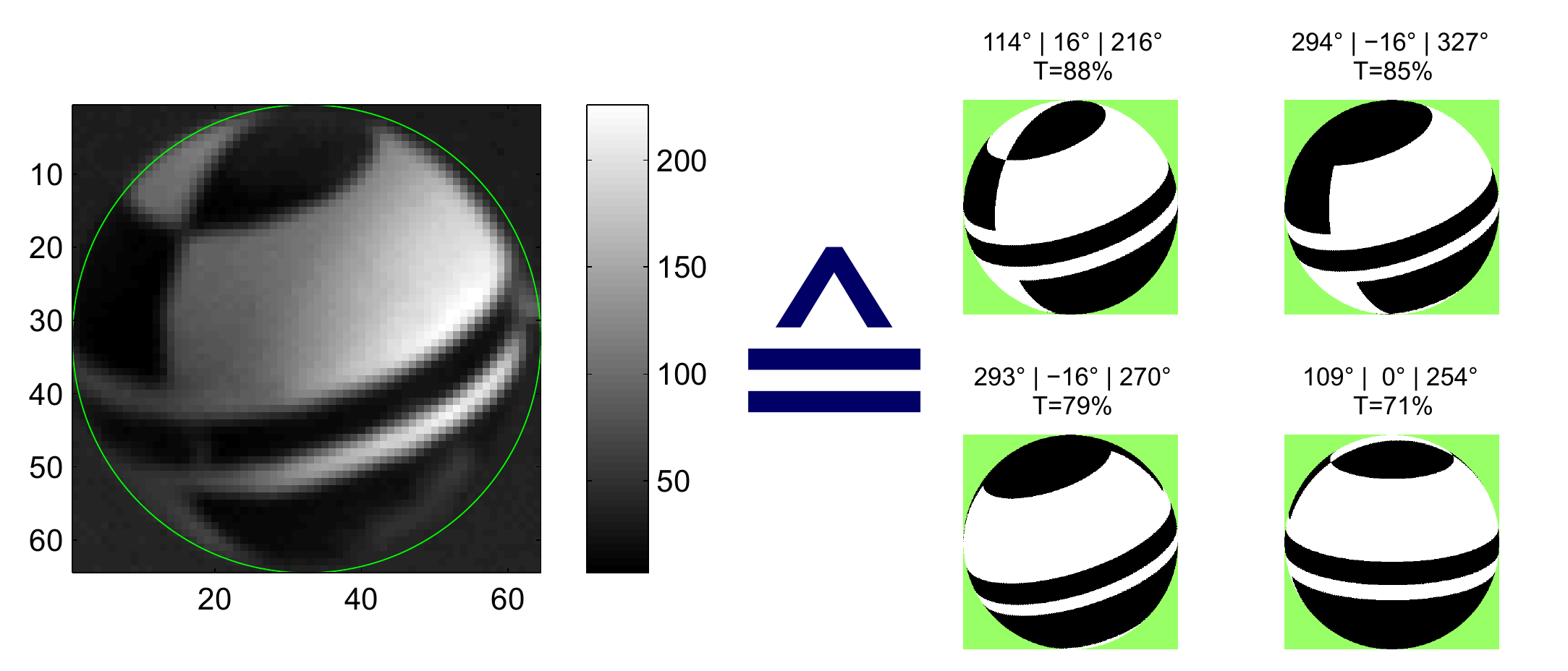} 
   \caption{Particle camera image (left) and corresponding candidates, after analysis of the possible orientations ({steps \emph{1a-c}} described in the text).}
   \label{fig:candidates}
\end{figure}

Experience shows that the identified POs usually cover several broad classes. They are thus separated into groups of images whose orientations differ by less than a rough threshold, approximately $30- 45\degree$. For each group, synthetic images are further added using a fine grid spacing, $\Delta_\text{fine}=3\degree$ (at this point `bad' images may cause the code to runaway; they are dropped and the code advances to the next time step).  The PO images are then rendered in real size and compared (using the projection $T$) to the image $\matrize I$. 
For each group, the code returns the final best guess, \ie the orientation with the maximum resamblance, thus drawing a list of \emph{candidates}, see \fig{fig:candidates} for an example of a particle with its corresponding candidates.

%

\subsubsection{Track Assembly}
After  identifying the candidates for each time step, the most likely orientation for each time step has to be determined. However, the candidate with the highest count of correct pixels is not  necessarily the best choice. Although counterintuitive, the direct use of 2 cameras seeing the particle at different angles does not simplify the problem, because in the case of a bad image, one camera  falsifies the choice of the candidates found by the other camera. Moreover, gimbal locks prevent the use of a predictor-corrector scheme for the prediction of the orientation. However, the norm of angular velocity is assumed to be smooth and we search the time series which  globally minimizes the sum  $\sum_{t} \xi(t)$ along the time series of the so called direct neighbor distance function:
\begin{equation}\begin{split}
\xi(t)\equiv\abs{ \vec\omega\big(\trip\theta(t),\trip\theta(t+\Delta t)\big) }=d\big( \trip\theta(t),\trip\theta(t+\Delta t) \big)\,\big/ \, \Delta t \ .
\end{split}\end{equation}
A direct neighbor  is the next valid time step at $t+\Delta t$. The distance between 2 orientations does not depend on the representation, ensuring the robustness of the algorithm even at gimbal locks. Minimizing $\sum_t \xi(t)$ is only meaningful for  small changes in orientation between two time steps, another requirement for  high (over)sampling rates.

Flow algorithms are highly efficient in finding a global optimum for a discrete set of candidates. The following is done for each camera without considering the extra information from the second camera. In a first step we remove all candidates with a resemblance $T<s_\text{quality}$ --  in practice $s_\text{quality}=0.5$.
Then a directed graph is built which  connects all candidates at time step $t$ with all their direct neighbors at the non-empty time step $t+\Delta t$. The cost function is chosen such that it takes into account both the change in orientation and the quality of the matching:
\begin{equation}
\begin{split}
C\big(\{\trip\theta_A,T_A\},\{\trip\theta_B,T_B\} \big)=d\big(\trip\theta_A,\trip\theta_B\big) ~\frac{2- T_A -T_B}{\Delta t} \ ,
\end{split}\label{eq:costfunc}
\end{equation}
with $\{\trip\theta_A,T_A\}$  a candidate at time $t$ and $\{\trip\theta_B,T_B\}$ a directly neighboring candidate at $t+\Delta t$.  

\begin{figure}[htb] 
   \centering
   \includegraphics[width=\columnwidth]{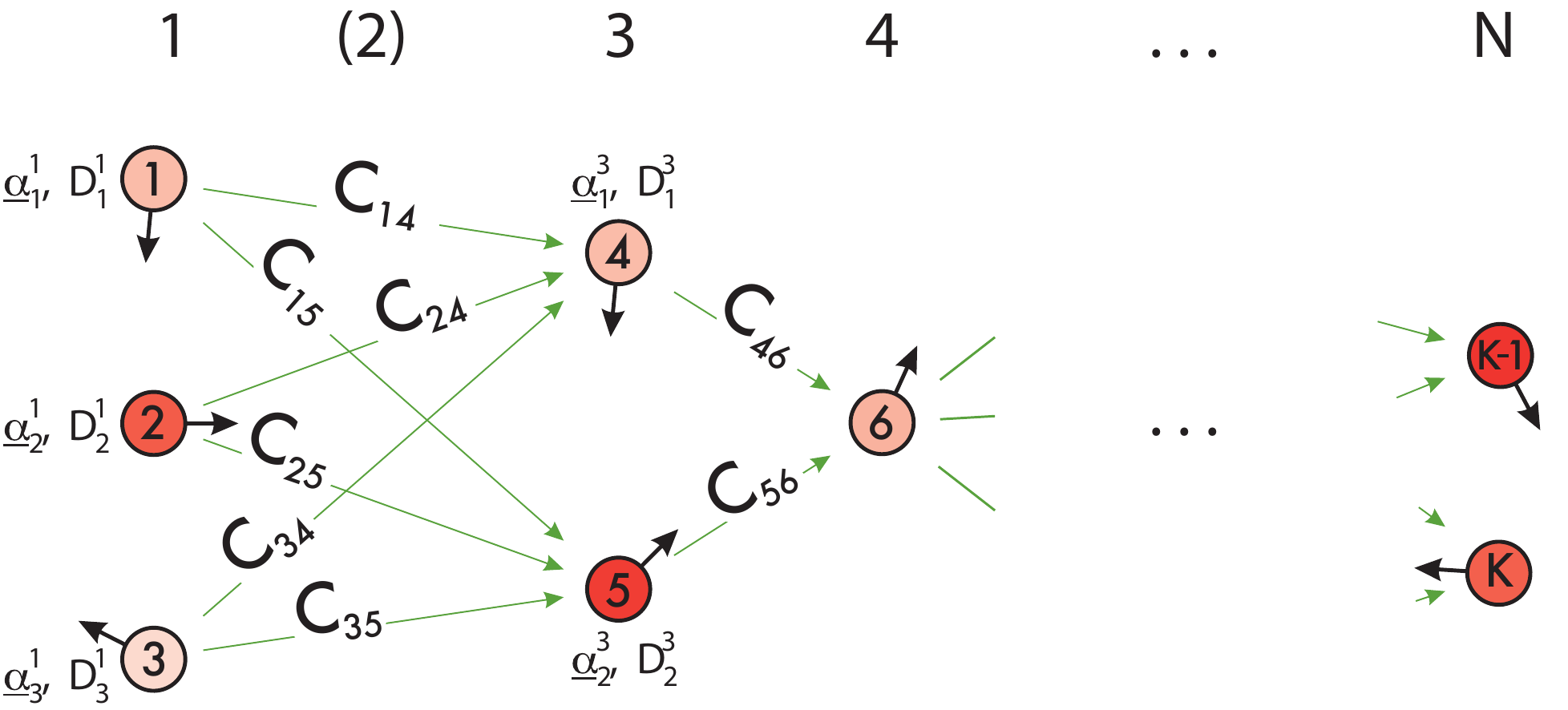} 
   \caption{Sketch of a graph connecting the possible candidates using the cost function  $C$ (cf.  \gleich{eq:costfunc}). In this example, no candidate could be identified at time setp 2. }
    \label{fig:candidates2}
\end{figure}

A Dijkstra path finding algorithm returns the sequence of candidates having a global minimum of the total cost, \ie the global minimum of change of orientation (weighted by the image quality) (cf. \fig{fig:candidates}). In most cases this algorithm  returns directly the time series of absolute orientation. Nevertheless, bad images introduce false candidates forcing the path finding algorithm to take a different, non-physical path. These points manifest as spikes in the direct neighbor distance function, $\xi(t)$. After a spike, there is no guarantee that the path is still physical. Therefore, we segment the time-series based on the  spikes. The second view (from the second camera) treated with the same algorithm contains the information to correct such wrong segments. From the camera calibration the rotation matrix which transfroms the orientations seen by one camera into the CS of the other one, is known. 
Therefore, both views are expressed in an intermediate, common CS where the segments with $d\big(\trip\theta_\text{cam1},\trip\theta_\text{cam2}\big)\gtrsim 30\degree$ can be corrected.  \\

The algorithm presented so far assumes an orthographic view. This condition holds only true if the particle center is on the optical axis of the camera or in the case one  uses tele-centric lenses. In the present experiment we do not, and the perspective effect alters the measured orientation (note that the parallax displacement corresponds to a change in the 2D projection, and hence to a rotation). The distortion induced by the perspective is characterized by the position of the particle center in the camera image, $\vec X$, and the focal length, $f$.  Common camera objectives allow only small angles, $\gamma_\text{persp}  \equiv \atan \left(\norm{\vec X} /f\right) \lesssim 15\degree$. As a consequence we assume that the shape of the particle does not change and we introduce an orientation matrix $R_\text{persp.}$ (taking advantage of the Rodrigues formula \gleich{eq:rodrigues}): 
\begin{equation}\begin{split}
\matrize R_\text{persp.}\big(\vec X=(x,y), f\big)=\matrize R\left( \frac{(-y,x,0)}{\sqrt{x^2+y^2}}, \atan \left(\frac{\norm{\vec X}} {f}\right)\right)
\end{split}\end{equation}
such that the measured orientation is related to the absolute orientation $\trip\theta_\text{abs}$ by $\matrize R\cong \matrize R_\text{persp.}~ \matrize R \big(\trip\theta_\text{abs}\big)$. The perspective distortion can then be removed from the orientation time series.\\

\begin{figure}[htb] %
   \centering
   \includegraphics[width=0.9\columnwidth]{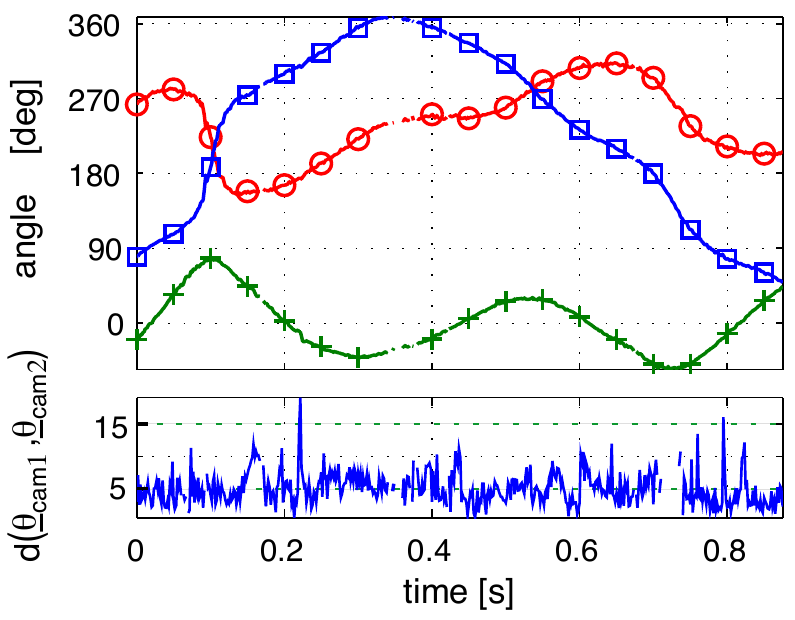} 
   \caption{A sample orientation track; it is $\RX=$ \Cred{ $\bigcirc$}, $\RY=$ \Cgreen{$+$}, $\RZ=$ \Cblue{$\square$}, the bottom plot shows the distance (in degrees) between the independent  orientation measurements from the 2 cameras.}
   \label{fig:sampleOTTrack}
\end{figure}

Finally, after correcting for perspective distortion,  a combined time-series of orientation can be built using the information from both views, if they are expressed in the same CS. Euler angles are not locally orthogonal, hence, we use the weighted mean of the orientation expressed in the axis-angle representation. The variance within a moving window of the  direct neighbor distance function, $\xi(t)$, proves to be a good error estimator of the noise, since for short times the particle is assumed to rotate smoothly. A sample orientation track is shown in the upper panel of \fig{fig:sampleOTTrack}.  

\subsection{Robustness}
A full study of the accuracy and robustness considering all possible distortions is beyond the scope of this article. In practice, the problems with real images are mainly caused by reflections, bad illumination, and objects  (such as bubbles or dirt particles) between the particle and the camera. The setup, light conditions and particle texture must be first tuned in order to optimize these parameters -- by trial and error methods. For the orientation algorithm \emph{per se}, we have used a series of synthetic images of known orientation. We found that the measurement error is 2\degree, which is smaller than the size, $\Delta_\text{fine}=3\degree$ of the fine grid used in the image processing (cf. paragraph~\ref{pp:cands}). A finer grid would improve the resolution for ideal images, but not for real images which, as stated above, always contain some amount of distortions or impurities. In addition, the fast dynamics of the particle and high frame rate ensure that wrong detection do not persist for longer than a few frames.  As a result, most defects are detected and skipped or interpolated or handled as part of post-processing (wrong orientations correspond to jumps in the direct neighbor distance function).

\begin{figure}[htb] %
   \centering
   \includegraphics[width=0.8\columnwidth]{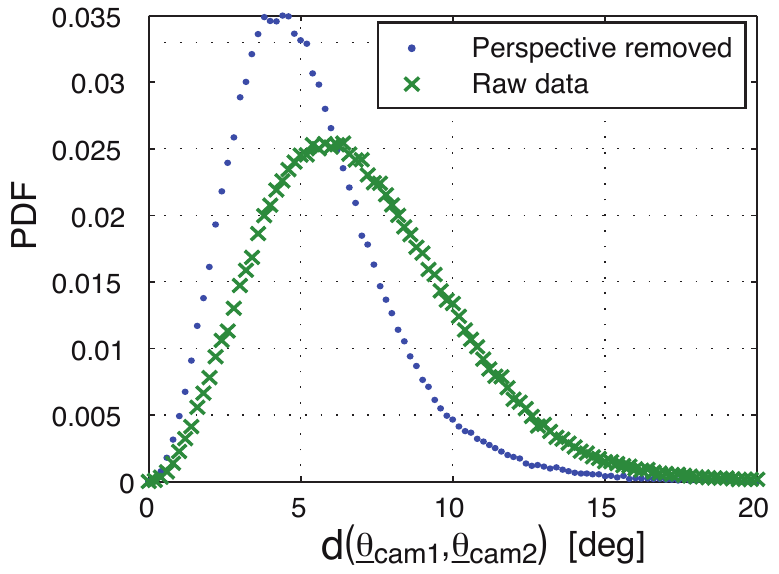} 
   \caption{Probability density function (PDF) of the distance between the orientations measured from cameras 1 and 2, without correction for perspective distortion ($\times$) and with it  ($\circ$).}
   \label{fig:dist2Cams}
\end{figure}

We illustrate the accuracy of the detection on 2 examples. The first one concerns the agreement between the orientation as estimated from each camera measurement. In the upper panel of \fig{fig:sampleOTTrack}, the combined 3 angles with respect to the Lab coordinate system are plotted. The lower panel shows the distance (in degrees of angle) between the two estimations, $d\left(\trip\theta_\text{cam1}, \trip\theta_\text{cam2}\right)$.  The probability density function (PDF) of these distances, computed with and without processing for perspective corrections are shown in \fig{fig:dist2Cams}. When the correction for perspective distortion is made, a weighted average leads to an absolute error equal to $3.5\degree$. 


%

\section{Results}\label{sec:physik}
The results in this section correspond to the flow created by counter-rotation of the driving disks at a rate of 3Hz. In this case the power injection is of the order of $\epsilon \sim 1.7$~W/kg, the integral time scale $T_L$ is about 0.3~s, so that the dissipative time and space scales are $\eta \sim 30\,\mu$m and $\tau_\eta \sim 1$~ms. As a result, the particle tracked has a size corresponding to $D/\eta \sim 600$ and $D/L_\text{int} \sim 0.6$ ($L_\text{int}$ is the scale at which energy is fed into the flow). The flow Reynolds number based on the Taylor micro-scale is $R_\lambda \sim 300$.  The camera frame-rate is 600~Hz, and the trajectories analyzed have been selected so that their duration is longer than $0.25~T_L$ and most range between $0.5$ and $3~T_L$.

\begin{table}[hb]
\caption{Characteristic values (mean~$\pm$~rms) for the particle motion. The angular variables are given for the lab and particle coordinate systems.  }
\label{tab:Stats}
\begin{tabular}{lc| r | r | r | r}
& & $x$~~  &  $y$~~ & $z$~~ & Norm~~ \\\hline
 $\vec v$ & $[m/s]$	&$0\pm0.28$ & $0 \pm 0.40$ & $0 \pm 0.37$ & $0.6 \pm 0.2$ \\
 $\vec a$ & $[m/s^2]$ & $-0.1 \pm 5.5$& $-0.3 \pm 5.6$&$0.2 \pm 6.0$&$8.4 \pm 5.3$\\\hline
$\vec\omega\lab$ &$[rad/s]$ & $-0.1\pm 8.0$ &   $0.2 \pm 7.8$ & $0.2 \pm 7.7$ & \multirow{2}{*}{$12.2 \pm 5.8$} \\
$\vec\omega\pa$ &$[rad/s]$ & $-0.1 \pm 7.1$ & $-0.3 \pm 8.2$ & $-0.1 \pm 8.1$ &  \\\hline
$\vec\alpha\lab$ &$[rad/s^2]$ & $-3 \pm 590$ &   $1 \pm 554$ & $0 \pm 559$ & \multirow{2}{*}{$820 \pm 530$} \\
$\vec\alpha\pa$ &$[rad/s^2]$ & $0 \pm 624$ & $2 \pm 516$ & $1 \pm 534$ &  \\
\end{tabular}
\end{table}%

\begin{figure}[tb] 
   \centering
   \includegraphics[width=0.7\columnwidth]{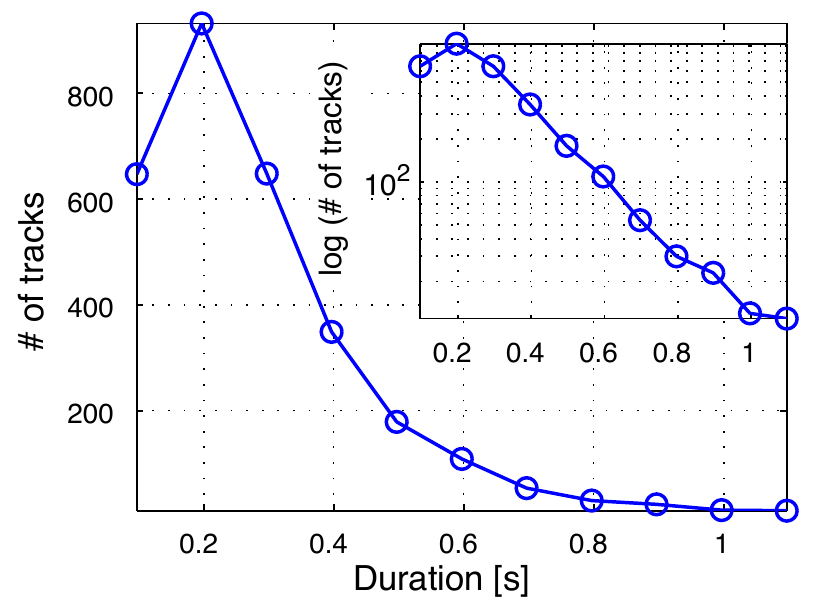} 
   \caption{Histogram of track segments. The exponential decay rate is of the order of the integral time $T_L$ of the flow.}
    \label{fig:seg}
\end{figure}

 \pagebreak[2]
\fig{fig:seg} shows a histogram of the duration of recorded tracks for which the 6D coordinates of the particle are recorded. It has an exponential tail (as it  was also the  case when using acoustic tracking~\cite{mordant:phd}). For very small times the histogram is biased by the fact than tracks shorter than 50 contiguous frames are discarded. Note also that long tracks are likely to correspond to trajectories spanning the flow volume, \ie a spatial extend over which the large scale (anisotropic) circulation cannot be ignored. \\

\begin{figure}[htb] 
   \centering
   \includegraphics[width=0.9\columnwidth]{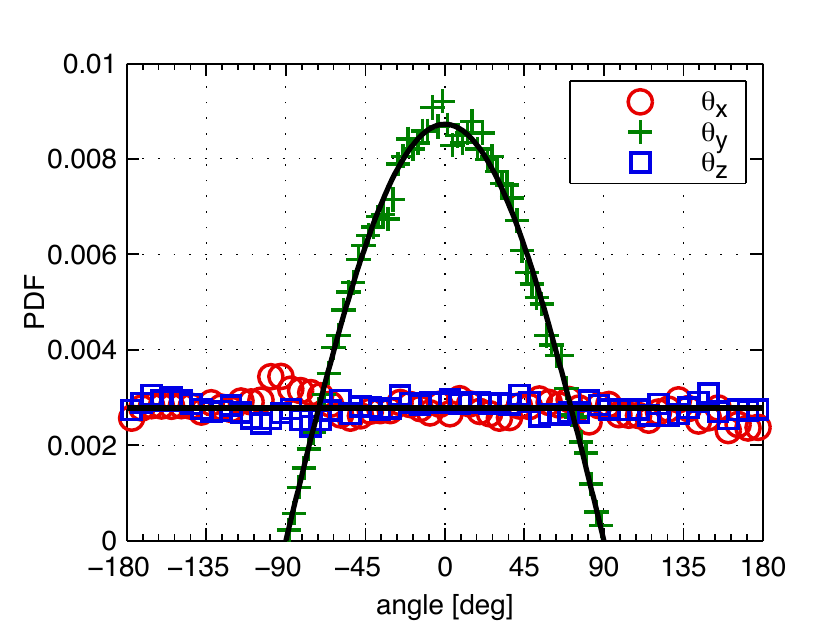} \\
   \caption{PDF of the orientation $\trip\theta=\{\RX,\RY,\RZ\}$, the solid lines correspond to a uniform sampling of the orientation space}
    \label{fig:PDFangle}
\end{figure}

However, one first result is that the particle explores uniformly the orientation space. This is seen in \fig{fig:PDFangle} showing the probability distribution functions of the Euler angles: as expected from a random distribution of orientations, the $\theta_x$ and $\theta_z$ components have a flat distributions spanning a $[-\pi, +\pi[$ interval, with the inner angle $\theta_y$ having a $cos(\theta_y)$ distribution over $[-\pi/2, \pi/2]$. 

\begin{figure}[htb] 
   \centering
   \includegraphics[width=0.8\columnwidth]{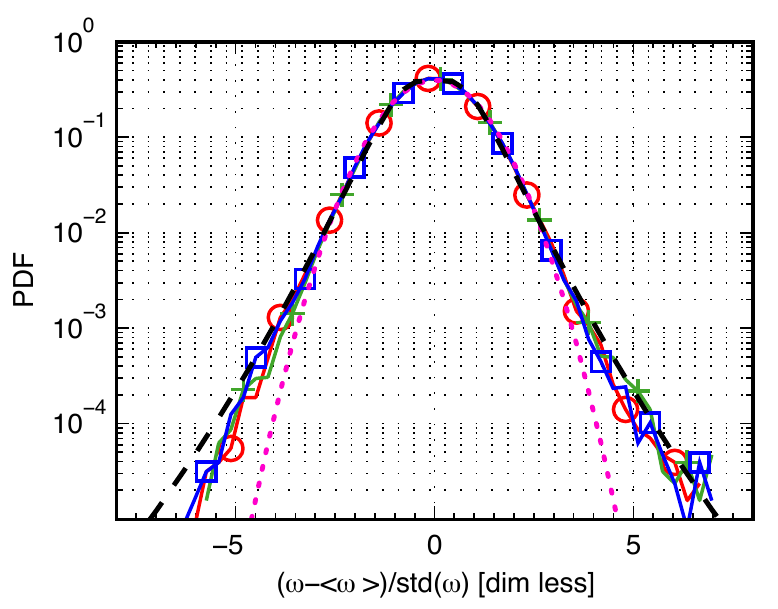} \\
   \caption{PDF of the (normalizd) components of the angular velocity,  $\omega_x=$ \Cred{ $\bigcirc$}, $\omega_y=$ \Cgreen{$+$}, $\omega_z=$ \Cblue{$\square$}, the dotted curve is a Gaussian and the dashed one shows a stretched exponential with $a=0.45~(F=4)$. }
    \label{fig:PDFangVel}
\end{figure}

Interesting features are observed for the rotation dynamics. The statistics of angular velocity fluctuations are shown in~\fig{fig:PDFangVel}. The distributions are symmetric. The three components, with respect to the Lab CS, follow the same statistics. This reflects the spherical symmetry of the particle; furthermore,  it also shows that the turbulent swirls at the scale of the particle have no preferred orientation. The mean of the angular velocity components (with respect to the Lab reference frame) is essentially zero, up to statistical error. The $rms$ amplitude of angular velocity fluctuation is of the order of 12~rad/s which is of the order of $u_\text{rms}/D = 30$~rad/s. That is, it corresponds to  the rotation that would result from imposing a velocity difference equal to  almost $u_\text{rms}$ across the diameter $D$ of the sphere. Note that it is also of the order of the rotation rate of the driving disks. 
The PDF themselves displays weakly stretched-exponential tails; for a quantitative estimation we use the fitting function:
\begin{equation}
{\Pi}_{a}(x) = \frac{e^{3a^2/2}}{4\sqrt{3}} \left( 1 - {\erf} \left( \frac{\ln \abs{x/\sqrt{3}} + 2 a^2}{a\sqrt{2}} \right)  \right) 
\end{equation}
which has been used extensively in the analysis of the intermittency of the translational motion of Lagrangian tracers~\cite{mordant:kernel} -- it stems from the approximation that the norm of the vector has a lognormal distribution. For the angular velocity, one finds a fitting parameter $a=0.45$, which corresponds to a flatness factor $F=4$. It would be $F=3$ for Gaussian statistics, so that our measurements show only a slightly non-Gaussian behavior for the angular velocity. This differs from the translational velocity, which is found to be slightly sub-Gaussian.\\

\begin{figure}[htb] 
   \centering
   \includegraphics[width=\columnwidth]{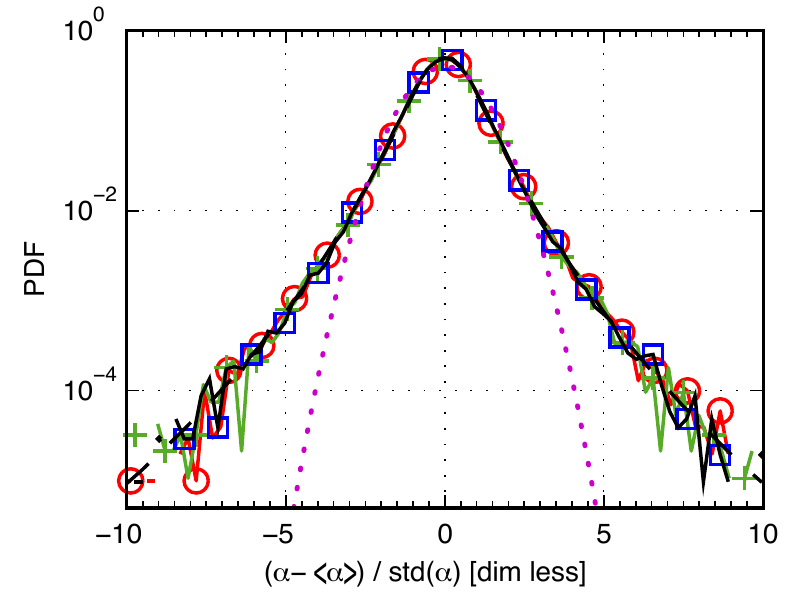} \\
   \caption{PDF of angular acceleration; it is $\vec\alpha_x=$ \Cred{ $\bigcirc$}, $\vec\alpha_y=$ \Cgreen{$+$}, $\vec\alpha_z=$ \Cblue{$\square$}, the dotted curve is a Gaussian and the dashed one shows a stretched exponential with $a=0.6~(F=7.6)$}
    \label{fig:PDFangAcc}
\end{figure}

The angular acceleration has  a strong non-Gaussian behavior, as seen in \fig{fig:PDFangAcc}. Again, the three components follow identical statistics: there is no preferred direction for the torques acting on the moving sphere (with respect to the Lab reference frame only -- the issue of lift forces is addressed elsewhere~\cite{liftforce}). The $rms$ amplitude of angular acceleration is about 800~rad/${\rm s}^2$, again of the order of $(u_\text{rms}/D)^2$. The statistics is strongly non Gaussian, a fit using the same stretched exponential distribution yields $a=0.6$, \ie a flatness factor $F \sim 7.6$. The angular acceleration can be viewed as an angular velocity increment over a very short time lag. Hence, the PDFs of angular velocity increments change shape with the length of the time lag -- from the one in \fig{fig:PDFangVel} for small time increments to the one in \fig{fig:PDFangAcc} for integral times. 

For comparison, we recall some features of the translational dynamics of the particle. It has statistical characteristics which are very close to the one reported for neutrally buoyant inertial particles with a size much closer to the dissipation scales of turbulence~\cite{mordant2001measurement,qureshi2008acceleration,brown:2009,volk2010dynamics}. The translational velocity follows a Gaussian distribution, its acceleration is strongly non-Gaussian, with stretched exponential tails. Using the stretched exponential distribution leads to $a=0.6$. One thus observes that  the angular variables have intermittent dynamics, just as the translational motion. The fact that it is quite pronounced, even for an object of  size close to the integral scale of motions  came as a  surprise and  deserves further investigations.

\section{Concluding remarks}
The focus of the work reported here has been to establish a technique for the study of angular and translational motion of a particle freely advected by a turbulent flow. We have shown that the measurement technique is robust, efficient, and accurate. As an application, we report here  the first observation of intermittency for the rotational dynamics of  an inertial particle. 

We note that the algorithm used to compute the angular velocity can be applied to a set of particle attached to a rigid body which are tracked using standard particle tracking algorithms. If one records the positions in space of 3 or more points, $\vec P_1 \dots \vec P_N$ at time $t$ and $t+\Delta t$, their motion can be split up into a translation of their center of mass (CM) plus a rotation. Once the translation part is subtracted, the rotation, $\matrize R_\text{kabsch}$, of the points $\vec P_1 \dots \vec P_N$ around their CM  can be computed efficiently using Kabsch's \cite{kabsch1976solution,kabsch1978discussion} algorithm. $\matrize R_\text{kabsch}$ is then the matrix representation of the change in orientation, and the angular velocity, $\vec\omega\pa$, (in the particle reference frame) at time $t$ can be extracted as done here. It should be pointed out that, one does not gain access to neither the angular velocity in the Lab reference frame, $\vec\omega\lab$, nor to the absolute orientation, $\trip\theta$.

The strong intermittency in the particle's rotation may eventually be traced back to the complex interaction between the particle and its wake. One notes that this is inherently  a finite size effect; for particles with very small diameters (compared to the Kolmogorov length) the translational and rotational dynamics are note coupled. For larger particles, as in our case, the influence of rotation on the motion of the particle is of interest, and will be the object of further analysis. One may also note that the influence of the inhomogeneity at large scale must be clarified.  Further measurements in a more isotropic turbulent flow (such as the Lagrangian Exploration Module\cite{RSI:LEM}) are underway.

\begin{acknowledgments}
We thank Aurore Naso for many fruitful discussions during the development and testing of the presented technique. This work was supported by ANR-07-BLAN-0155, and by PPF `Particules en Turbulence' from the Universit\'e de Lyon.
\end{acknowledgments}


\bibliography{biblioRSIOT}

\end{document}